\documentclass[conference]{IEEEtran}
\IEEEoverridecommandlockouts
\usepackage{cite}
\usepackage{amsmath,amssymb,amsfonts}
\usepackage{algorithmic}
\usepackage{graphicx}
\usepackage{textcomp}
\usepackage{xcolor}
\def\BibTeX{{\rm B\kern-.05em{\sc i\kern-.025em b}\kern-.08em
    T\kern-.1667em\lower.7ex\hbox{E}\kern-.125emX}}

\usepackage{cite}
\usepackage{amsmath,amssymb,amsfonts}
\usepackage{graphicx}
\usepackage{textcomp}
\usepackage{xcolor}
\usepackage[hidelinks]{hyperref}

\begin{document}

\title{Transforming Mentorship: An AI Powered Chatbot Approach to
University Guidance\\

}

\author{\IEEEauthorblockN{Mashrur Rahman}
\IEEEauthorblockA{\textit{Computer Science and Engineering} \\
\textit{Brac University}\\
Dhaka, Bangladesh \\
mashrur.rahman@g.bracu.ac.bd}
\and
\IEEEauthorblockN{Mantaqa abedin}
\IEEEauthorblockA{\textit{Computer Science and Engineering} \\
\textit{Brac University}\\
Dhaka, Bangladesh\\
mantaqa.abedin@g.bracu.ac.bd}
\and
\IEEEauthorblockN{Monowar Zamil Abir}
\IEEEauthorblockA{\textit{Computer Science and Engineering} \\
\textit{Brac University}\\
Dhaka, Bangladesh \\
monowar.zamil.abir@g.bracu.ac.bd}
\and
\IEEEauthorblockN{MD. Faizul Islam Ansari}
\IEEEauthorblockA{\textit{Computer Science and Engineering} \\
\textit{Brac University}\\
Dhaka, Bangladesh \\
faizul.islam.ansari@g.bracu.ac.bd}
\and
\IEEEauthorblockN{Adib Reza}
\IEEEauthorblockA{\textit{Computer Science and Engineering} \\
\textit{Brac University}\\
Dhaka, Bangladesh \\
adib.reza@g.bracu.ac.bd}
\and

\IEEEauthorblockN{Dr. Farig Yousuf Sadeque}
Associate Professor
\IEEEauthorblockA{\textit{Computer Science and Engineering} \\
\textit{Brac University}\\
Dhaka, Bangladesh \\
farig.sadeque@bracu.ac.bd}
\and
\IEEEauthorblockN{Niloy Farhan}
Adjunct Lecturer
\IEEEauthorblockA{\textit{Computer Science and Engineering} \\
\textit{Brac University}\\
Dhaka, Bangladesh\\
ext.niloy.farhan@bracu.ac.bd}
}

\maketitle

\begin{abstract}
University students face immense challenges during their undergraduate lives, often being deprived of personalized on-demand guidance that mentors fail to provide at scale. Digital tools exist, but there is a serious lack of customized coaching for newcomers. This paper presents an AI-powered chatbot that will serve as a mentor for the students of BRAC University. The main component is a data ingestion pipeline that efficiently processes and updates information from diverse sources, such as CSV files and university webpages. The chatbot retrieves information through a hybrid approach, combining BM25 lexical ranking with ChromaDB semantic retrieval, and uses a Large Language Model,  LLaMA-3.3-70B, to generate conversational responses. The generated text was found to be semantically highly relevant, with a BERTScore of 0.831 and a METEOR score of 0.809. The data pipeline was also very efficient, taking 106.82 seconds for updates, compared to 368.62 seconds for new data. This chatbot will be able to help students by responding to their queries, helping them to get a better understanding of university life, and assisting them to plan better routines for their semester in the open-credit university.
\end{abstract}

\begin{IEEEkeywords}
Chatbot, Corpus-based, Natural Language Processing (NLP), Per-
sonalized guidance, Interactive chatbot, Educational technology, Informal dialogue, Information retrieval, Open-credit universities, Bias assessment
\end{IEEEkeywords}

\section{Introduction}
Due to the dynamic academic environment, large number of students with fewer faculties and staffs, and difficult university program policies and procedures, challenges were present throughout the four years of university education. Open credit universities face challenges in obtaining accurate policy information, selecting appropriate courses, scheduling classes, and managing limited time with mentors due to mentor shortages. Technology has given students many resources, but on-demand and personal help is still lacking. This is especially risky for first-year students who sometimes struggle with the new environment and may need additional guidance. To fill this gap, we will provide a corpus-based chatbot that also serves as a student companion. Chatbots offer many benefits, including: providing fast, personalized help (Shawar Atwell, 2007)\cite{Shawar2007}. Digital agents can understand natural languages and initiate HHI-like conversations with users. Through NLP, the chatbot can better understand student questions and answers, providing necessary assistance and guidance (Brandtzaeg Følstad, 2017)\cite{Brandtzaeg2017}.Implementing chatbots in education is now mandatory worldwide. The bot can quickly respond to basic questions and queries, saving time and energy for faculty and staff (Winkler S\"{o}llner, 2018)\cite{Winkler2018}. Students can ask the chatbot questions about current topics, schedule class times, and get help with general academic matters. This automation enables algorithms to solve tasks automatically, freeing teachers and mentors to focus on non-automated tasks like student behavior analysis, lesson attention, and learning efficiency (Folstad Skjuve, 2019)\cite{Folstad2019}. After the discussion, we will design and build an efficient chatbot to help first-year students with study difficulties. The proposed chatbot aims to improve student learning by providing timely information about semesters and course topics. It will also benefit those giving advice, relieve faculty pressure, and improve advising efficiency. Thus, using a specially designed corpuschat as part of the digital mentoring system to solve university information technology challenges is feasible. It can reduce the gap for individual attention and help deliver education technologies to support individuals. To conclude, this research aims to create a practical tool that enhances student learning and reduces faculty and staff workload, enhancing overall effectiveness and efficiency.

\section{RELATED WORKS}

\subsection{Backround and Development}
The development of chatbots has evolved significantly over time, starting with ELIZA, the first chatbot developed in 1966, which used pattern matching and response templates but lacked depth in conversations (Adamopoulou and Moussiades, 2024) \cite{Adamopoulou2021}. Over time, chatbot technology has improved, with modern models capable of maintaining context-aware and dynamic interactions through advanced machine learning algorithms.

Suta et al. (2020) \cite{Suta2020} highlight the challenges in designing intelligent chatbots that can accurately interpret and respond to human language. Despite advances in natural language processing (NLP) and machine learning (ML), chatbots still struggle with understanding user queries contextually. Similarly, Ranoliya, Raghuwanshi, and Singh (2017) \cite{Ranoliya2017} investigate chatbot applications in education, leveraging artificial intelligence markup language (AIML) and latent semantic analysis (LSA) to enhance response accuracy.

Nguyen (2019) \cite{Nguyen2019} examines the role of question-answering (QA) systems in the biomedical field, emphasizing the importance of robust information retrieval methods. Despite using TF-IDF ranking and entity extraction tools such as MetaMap, these systems still struggle with complex medical terminologies and phrase ambiguities. The limitations in semantic understanding highlight the necessity of using deep learning models for improved comprehension and contextual relevance.
\subsection{Chatbots in Education}
Hwang and Chang (2021) \cite{Hwang2021} investigate the opportunities and challenges of educational chatbots, recognizing their role in enhancing learning experiences. These chatbots facilitate interactive, real-time assistance for students, improving communication and providing individualized feedback. However, the lack of comprehensive research on chatbot effectiveness in diverse educational environments remains a challenge.

El-Ashmawi et al. (2023) \cite{ElAshmawi2023} develop a chatbot for university-related queries using AIML and the RASA framework, demonstrating how AI-driven bots enhance student support. Formica et al. (2023) \cite{Formica2023} explore the use of SPARQL queries for knowledge-based chatbot systems, allowing chatbots to fetch structured data from knowledge bases like Wikidata and DBpedia.

Nguyen et al. (2023) \cite{Nguyen2023} address the issue of ‘summer melt,’ where a large percentage of students fail to complete the enrollment process despite being accepted into college. Their chatbot, Lilo, assists students with advising and social-emotional support to improve retention rates. Similarly, Dinh and Tran (2023) \cite{Dinh2023} emphasize the role of chatbots in automating services across domains such as education, healthcare, and e-commerce.
\subsection{Chatbot Architecture}
In customer service automation, Pandya and Holia (2023) \cite{Pandya2023} introduce Sahaay, a chatbot framework utilizing LangChain and large language models (LLMs) for contextual and personalized assistance. The improved performance of chatbots using Flan T5 XXL models showcases their application in diverse sectors. Shahriar et al. (2023) \cite{Shahriar2023} focus on Bengali Question-Answer Generation (QAG), addressing the scarcity of QA datasets in low-resource languages.

Cui et al. (2017) \cite{Cui2017} propose SuperAgent, a customer service chatbot for e-commerce, demonstrating how vast data sources can be leveraged for improved user interactions. This chatbot retrieves information from product descriptions, customer reviews, and FAQs to generate relevant responses, reducing human intervention in routine queries. Meanwhile, Sweidan et al. (2021) \cite{Sweidan2021} design an Android-based university assistant chatbot for the University of Jordan, featuring bilingual capabilities for academic inquiries.

Swanson et al. (2019) \cite{Swanson2019} introduce recurrent neural networks (RNNs) in conversational retrieval chatbot models to improve response relevance. This approach addresses the issue of context retention in chatbots, ensuring more accurate responses. Similarly, Atmauswan and Abdullahi (2022) \cite{Atmauswan2022} discuss the transition from early chatbots like ELIZA and ALICE to modern AI-driven chatbot systems, which leverage NLP and deep learning for enhanced performance.

Mukherjee et al. (2023) \cite{Mukherjee2023} tackle the challenge of generating polite responses in chatbot interactions, proposing a three-step model for training chatbots to respond with appropriate politeness markers. Their work highlights the importance of ensuring natural and courteous AI conversations. Lee et al. (2023) \cite{Lee2023} develop PEEP-Talk, a chatbot designed for English language learning, leveraging real-world situational dialogues for more effective language acquisition.

Campese et al. (2024) \cite{Campese2024} discuss advancements in question-answering (QA) retrieval models, proposing an unsupervised pre-training (PT) method for ranking semantically matched questions. Their work improves response selection in chatbot systems for FAQs and forum discussions.

\section{METHODOLOGY}
\subsection{Dataset Description and Collection}
The question-answer corpus for training the Question-Answering Language Model on BRAC University students was created for this study. Students' topics and questions are covered by 1866 data points. The goal was to answer all student questions from the first day to the last, including thesis/internship questions. All data is in English. Data on students' daily academic questions was collected from various sources. The data was carefully selected from three primary sources:\vspace{0.1cm} 

\begin{figure}[htbp]
    \centering
    \includegraphics[width=0.4\textwidth]{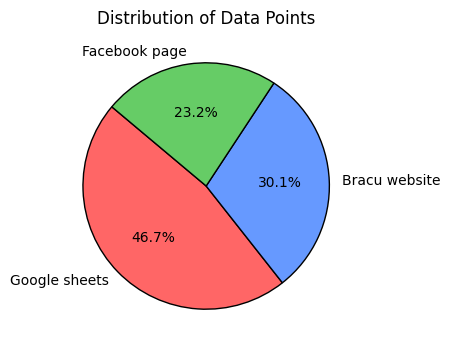} 
    \caption{Distribution of Dataset.}
    \label{fig:Distribution of Dataset}
\end{figure}

\textbf{Google Sheets:} BRACU provides a google sheet for CSE routines each semester. There are different section-wise course routines, including faculty, class day and hour, class room number, etc. That sheet also lists faculty and coordinators. Students need these data points daily. Data points from course sections totaled 529. Our corpus had 232 faculty and 76 coordinator data points. Another sheet had 35 course prerequisite data points.\vspace{0.1cm} 

\textbf{BRAC University website:} This website was used to gather information on campus life, learning facilities, and student services. As an example:

\textcolor{red}{"Question"}: \textcolor{black}{"What is the grading scale utilized by BRAC University?"}, \\

\textcolor{red}{"Category"}: \textcolor{black}{"Exams and Grading"}, \\

\textcolor{red}{"Answer"}: \textcolor{black}{"BRAC University uses a grading scale based on the Grade Point Average (GPA) system, where letter grades correspond to specific grade points. The typical grading scale is as follows: A: 4.0 (Excellent), A-: 3.7, B+: 3.3, B: 3.0 (Good), B-: 2.7, C+: 2.3, C: 2.0 (Satisfactory), C-: 1.7, D+: 1.3, D: 1.0 (Pass), F: 0.0 (Fail)"}.

\textbf{Facebook groups:} Student and alumni forums that discuss university experiences were used to identify frequently asked and answered questions. These groups provided useful information on student issues and daily life. Our corpus gained 432 data points from this method. \vspace{0.1cm}

The Facebook group dataset is divided into 14 categories that reflect major student learning areas and program services. Advice, academic progress, exams grading, financial aid payment, library, international student, technical support, policies procedures, TARC, campus, extracurricular activities, admission, transportation, and thesis internship are examples. All categories should match students' interests, making the model applicable to any aspect of university life. \vspace{0.1cm} 

The “campus” category receives nearly 400 questions and is the most popular topic among students for campus location and facilities. Popular topics include ‘advising’ and ‘payment financial aid scholarship’, indicating that students need information on academic advising and scholarships and financial aid. Other subtopics include “academic process info”, “TARC”, and “thesis and internship”, confirming that students often seek information about academic processes, TARC campuses, and internship and thesis requirements. Libraries, extracurricular activities, and international students are further down the chart with fewer questions. They may have fewer questions because they are niche or fewer students are interested.
\vspace{0.1cm} 

\begin{figure}[htbp]
    \centering
    \includegraphics[width=0.45\textwidth]{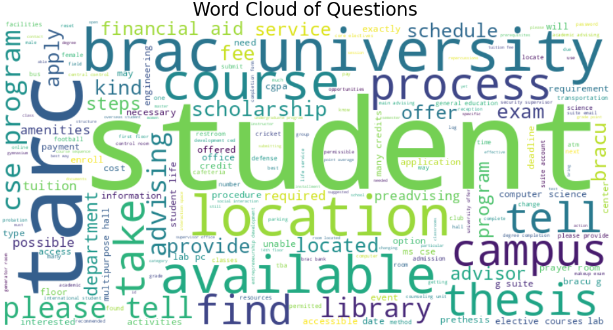} 
    \caption{Word Cloud of Questions.}
    \label{fig:Word Cloud of Questions}
\end{figure}

The dataset's most common terms are shown in Figure 2. The size of each word indicates its frequency in the questions. Common terms such as ‘student,’ ‘tarc,’ ‘university,’ ‘course,’ ‘location,’ and ‘thesis’ indicate that many questions focus on student issues, residential campus, course details, and university processes. Using terms like “financial aid,” “scholarship,” “library,” and “process” suggests students seek information on procedures, facilities, and funding.
Our dataset contains static and dynamic questions. Static questions usually have fixed answers. Total credits needed for degree completion are not likely to change frequently. Different answers are given to dynamic questions each semester, such as academic schedule and offered electives. In this thesis phase, we combined two types of questions for thorough analysis.
Questions and answers vary in length and difficulty, with some sharing factual information and
others offering post-experience advice. Questions can range from basic factual ones, such as “what eating places are there in TARC?” Descriptive to exploratory problem-solving questions, such as “How should students allocate their time at BRACU?” Answers should be brief but detailed, providing directions or descriptions relevant to the question at hand. \vspace{0.1cm} 

\subsection{Data Preprocessing}

Steps in data preprocessing include cleaning, tokenizing, lemmatizing, and transforming questions and answers. Since our dataset consists of question-answer pairs, we aim to standardize and prepare it for machine learning models. We designed the workflow to minimize inconsistencies and noise in raw textual data. \vspace{0.1cm} 

1. \textbf{Dropping Nulls:} \vspace{0.1cm} 
First, we loaded our datasets into Pandas DataFrame to remove unnecessary and null data. This made changes and analysis easier. We looked for null values in the question-answer pairs from the Facebook group. We eliminated missing values such as “questions” and “answers”. Thus, data entry was complete. After removing null values, 423 question-answer pairs remained out of 432. Incomplete data could lead to biases and errors during model training, making data cleaning crucial. \vspace{0.1cm} 

2. \textbf{Lowercasing and Punctuation Removal:} After removing nulls, we normalized text. For normalization, we converted all text in the “questions” and “category” columns to lowercase to avoid different tokens for the same word due to capitalization variations. Applying the same context to “Chatbot” and “chatbot” is an example. We removed punctuation marks as they were not necessary for our tasks and could lead to inconsistencies if left unprocessed. \vspace{0.1cm}

3. \textbf{Tokenization:} Tokenization split questions into word tokens. Tokenization simplifies text analysis by breaking it down into smaller components. Also crucial to Natural Language Processing. We transcribed each question into a list of words for further processing. \vspace{0.1cm}

4. \textbf{Lemmatization:} Tokenization was followed by WordNetLemmatizer. Used for lemmatization. Lemmatization helps maintain consistency across word variations by restoring their original form. If the dataset contains words like “Adding” and “Added”, they will be reduced to “Add”. We used this step to reduce vocabulary size and ensure semantically similar words are handled similarly in machine learning models.

5.\textbf{Keras Tokenizer Numerical Transformation:} After tokenization and lemmatization, we began numerical transformation. This method maps question words to unique integer values. Machine learning models require numerical values for training, not text-based data processing. Thus, we numericalize them.After fitting each word to the text, the tokenizer assigns an integer. Later, it turns all text into numbers. The model could use these sequences as inputs. \vspace{0.1cm} 

6. \textbf{Sequence Padding:} For machine learning models, input data must be of the same length. For that, sequences are padded to 50 words. Pads ensure all input sequences have the same number of elements.
Padding maintains the same number of elements, streamlines batch processing, and prevents errors during model training. When a sequence has fewer than 50 words, zeros are added to meet the mandatory length. Standardizing and analyzing inputs is also ensured. \vspace{0.1cm}

7. \textbf{RecursiveCharacterTextSplitter:} The RecursiveCharacterTextSplitter is a tool for breaking down large files into manageable chunks, such as characters, paragraphs, or sentences. Our system divides documents into 1000 characters with 200 character overlap, allowing for context information from previous chunks. This preprocessing step is crucial as large documents can hinder model retrieval. It simplifies text storage in vector databases by allowing segmentation, retrieval, and processing in smaller increments, significantly improving the retrieval and generation pipeline. \vspace{0.1cm} 

8. \textbf{Google Generative AI Embedding:} The Google Generative AI Embeddings model embeds PDF and other document text. These quantized representations of text meaning and context allow the system to discover "similarity" between texts. This procedure helps the system scan by identifying the best document fragments to answer user queries. The model provides only the most relevant information for a response using these embeddings.\vspace{0.1cm} 

Why Use It for BRAC University Chatbot? \vspace{0.1cm} 

Efficient Document Retrieval: The university handles many documents, such as course catalogs and policies, making it easier to retrieve them when students have questions. \vspace{0.1cm} 

Contextual Relevance: Embeddings help the chatbot find relevant data from university sources, ensuring accurate responses. \vspace{0.1cm}

9. \textbf{all-MiniLM-L6-v2 sentence-transformers:} The transformers-based model sentence-transformers/all-MiniLM-L6-v2 creates vector embeddings from text. This model effectively generates embeddings for identifying semantic similarities in diverse sentence structures. Create embeddings for user queries and datasets to utilize semantic similarity during retrieval. \vspace{0.1cm} 

10. \textbf{Facebook AI Similarity Search:}  
FAISS stores document chunks as vector embeddings for similarity search. After text segmentation, Google Generative AI Embeddings generate vector embeddings. FAISS uses these embeddings to create an index for efficient approximate nearest neighbor (ANN) search. After a user enters a question, FAISS finds a similar vector embedding and extracts the most semantically related chunks from documents. FAISS is essential for efficient retrieval in large datasets, as chunks are fed to the model to produce an answer. \vspace{0.1cm} 

11. \textbf{Chroma DB:} Chroma DB is a vector database which can store vector embedding of large dimensions. It is similar to FAISS but it can store metadata and it can dynamically update embeddings. This metadata can be sent to the LLM to provide rich context and enhance relevant answer generation.\vspace{0.1cm} 

\subsection{Data Ingestion Pipeline}\label{AA}
This section talks about the design and implementation of the data ingestion pipeline
which is used to extract data from multiple CSV files and webpages. Finally, load
them into a relational database and eventually into a vector database.\vspace{0.1cm}

The purpose of this data ingestion pipeline is two-fold: \vspace{0.1cm}

\textbf{Structured Data Storage:} A relational database provides structured data storage for raw tabular information including questions and answers along with course
schedules and faculty data and many more. It is specially useful to support database
updates and modifications and maintain auditability. \vspace{0.1cm}

\textbf{Semantic Search and Retrieval:} Repurposing the stored records with vector
embeddings creates possibilities for Chroma to provide advanced searching and
question-answering capabilities. \vspace{0.1cm}

The system achieves compatibility with both standard SQL database queries along
with complex natural-language queries by dividing database management from vector embedding operations.\vspace{0.1cm}

For CSV files, the SQLite database creation step involves creating tables for each file. A CSV file containing faculty list information has 6 columns (Initial, Name, Designation, Status, Room, and Email), while a CSV file containing course prerequisite information has 3 columns (Course, Pre-Requisite, and Full Chain Pre-Requisite), resulting in different schemas for each table.
When populating tables, each row stores time to avoid repeating data ingestion. Data ingestion timestamps are automatically added to new rows during record insertion. Adding timestamps to records is crucial for identifying new and updated records in later stages. \vspace{0.1cm}

\begin{table}[htbp]
\centering
\scriptsize 
\begin{tabular}{|c|p{1.2cm}|p{1.5cm}|p{2cm}|p{1cm}|}
\hline
\textbf{id} & \textbf{Course} & \textbf{Prerequisite} & \textbf{Full Chain Prerequisite} & \textbf{Timestamp} \\ \hline

1  & CSE111 & CSE110 (HP) & CSE111--CSE110 & 2025-01-26 14:32:30 \\ \hline
2  & CSE220 & CSE111 (HP), CSE230 (HP) & CSE230--CSE111--CSE110 & 2025-01-26 14:32:30 \\ \hline
3  & CSE221 & CSE220 (HP) & CSE220--CSE111--CSE110 & 2025-01-26 14:32:30 \\ \hline
4  & CSE250 & PHY112 (SP) & None & 2025-01-26 14:32:30 \\ \hline
5  & CSE251 & CSE250 (HP) & CSE250 & 2025-01-26 14:32:30 \\ \hline
6  & CSE260 & CSE251 (HP) & CSE251--CSE250 & 2025-01-26 14:32:30 \\ \hline
7  & CSE310 & CSE370 (HP) & CSE370--CSE221--CSE220--CSE111--CSE110 & 2025-01-26 14:32:30 \\ \hline
8  & CSE321 & CSE221 (HP) & CSE221--CSE220--CSE111--CSE110 & 2025-01-26 14:32:30 \\ \hline
9  & CSE330 & MAT216 (HP) & MAT120--MAT110 & 2025-01-26 14:32:30 \\ \hline
10 & CSE331 & CSE221 (HP) & CSE221--CSE220--CSE111--CSE110 & 2025-01-26 14:32:30 \\ \hline

\end{tabular}
\caption{Example of populated table in SQLite dataset}
\label{tab:prerequisites}
\end{table}

\vspace{0.1cm}

\textbf{Webpages:} The system starts by doing URL retrieval from a text file while validating file formats. The system uses trafilatura~\cite{trafilatura} to parse HTML responses and extract text content by discarding scripts and styles from each URL request that happens asynchronously. SHA-256~\cite{nist_sha} hashing acts as the tracking mechanism for modified extracted content. The system performs hash comparison against its stored records to check for updates. The system does not require a new entry creation once it determines that the content remains unchanged to avoid duplication. A new entry is created while maintaining historical version data for analysis. The program utilizes a local SQLite database to maintain scraped material and linked URLs to supply quick retrieval and a structured file system. Every version of web page content has its own individual storage space which enables future access for analysis along with maintaining a versioned dataset. A data-ingestion script requires execution through an alert to process newly inserted content before it can be indexed embedded or analyzed semantically. The combined method produces data acquisition which seamlessly integrates change identification with effective storage delivering uninterrupted monitoring along with information renewal. The Figure 3 shows the whole workflow of fetching website data.\vspace{0.1cm}

\begin{figure}[htbp]
    \centering
    \includegraphics[width=0.4\textwidth]{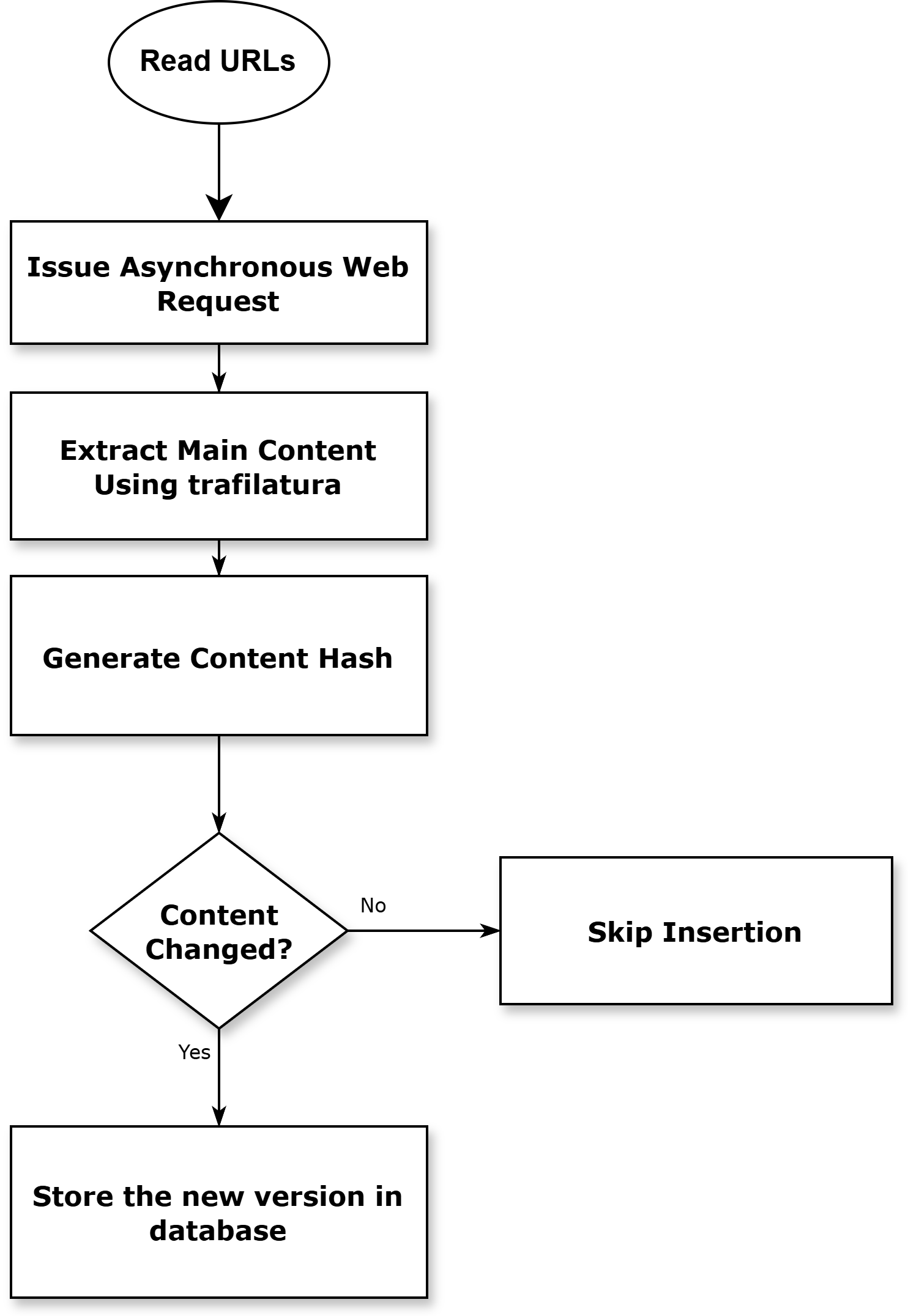} 
    \caption{Workflow of fetching webpages.}
    \label{fig:Workflow of fetching webpages.}
\end{figure}

\textbf{Extracting from database and store in vector database} \vspace{0.1cm}

Ingestion into the knowledgebase follows database population. Chroma DB is our vector database. Include steps: \vspace{0.1cm} 

\textbf{Database Access:} SQLite opens a new connection to the same database. Database processes choose fields from each table. The EnglishQA table automatically retrieves the Question, Answer, and Timestamp fields with each query. \vspace{0.1cm}

\textbf{Increasing Consumption Timestamping:}The script keeps a dictionary of timestamp values for each table. The pipeline retrieves data points from previously ingested tables when the Timestamp value exceeds the last threshold. Otherwise, all rows are consumed. The optimization system eliminates duplicate records and allows for incremental updates. \vspace{0.1cm}

\textbf{Text Generation from Records and Chunking:} The script generates entry-specific text summaries from records and chunking. Vector embedding models use text. The system generates several short documents from each document entry to store faculty, scheduling, and email data. Using smaller segments improves embedding granularity, resulting in improved search accuracy for specific queries.Any large document is chunked using RecursiveCharacterTextSplitter. Large documents are memory-intensive, so chunking speeds up processing and improves retrieval. \vspace{0.1cm}

\textbf{Embed Each Document:} The embedding model uses custom SentenceTransformer embedding methods to embed each textual segment. Text is converted to numerical vectors by sequential processing. Semantic consistency in embeddings allows similarity computation.\vspace{0.1cm}

\textbf{Storing in Chroma (Vector DB):} Chroma (Vector DB) stores vector embeddings and metadata labels like document ID and table names. After pre-insertion checks for documents with identical IDs, the ingestion script deletes outdated entries and updates them. Each unique record has one vector database data representation. \vspace{0.1cm}
\begin{figure}[htbp]
    \centering
    \includegraphics[width=0.5\textwidth]{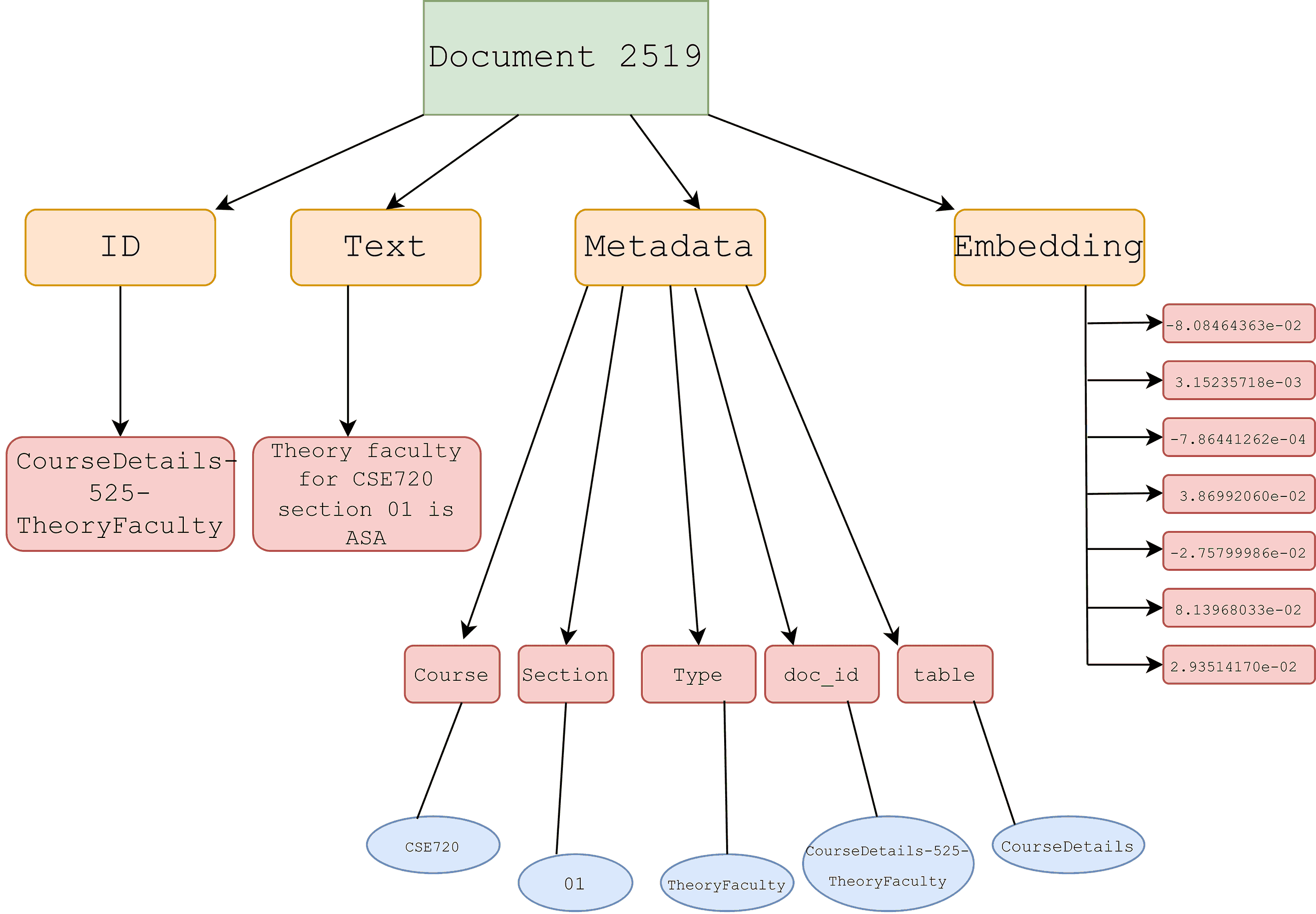} 
    \caption{Representation of unique record stored in Chroma DB.}
    \label{fig:Representation of unique record stored in Chroma DB.}
\end{figure}

\textbf{Advantages of this Data Ingestion Pipeline} \vspace{0.1cm}

\textbf{Incremental Updates:} The pipeline takes timestamp updates to limit its data consumption only to new or modified data that decreases computation time and prevents repeated embedding generation. \vspace{0.1cm}

\textbf{Granular Embedding:} Data splitting at the record level results in improved search
precision through separated textual segments. Users can input demands regarding
record subsections (example: ”theory faculty for CSE101”) to obtain the system’s
most suitable data segments. \vspace{0.1cm}

\textbf{Maintaining a Single Source of Truth:} The SQLite database receives data from
users through validated structures yet users benefit from Chroma’s flexible semantic
search system which omits repetitive detail storage.
Scalability: Extra CSV files or new tables merge with the current system after
straightforward adjustments to processing. The vector database retains its fast
retrieval speeds after scaling or sharding operations as your data volume grows.\vspace{0.1cm}

\

The figure 5 shows the overall workflow of database population

\begin{figure}[htbp]
    \centering
    \includegraphics[width=0.37\textwidth]{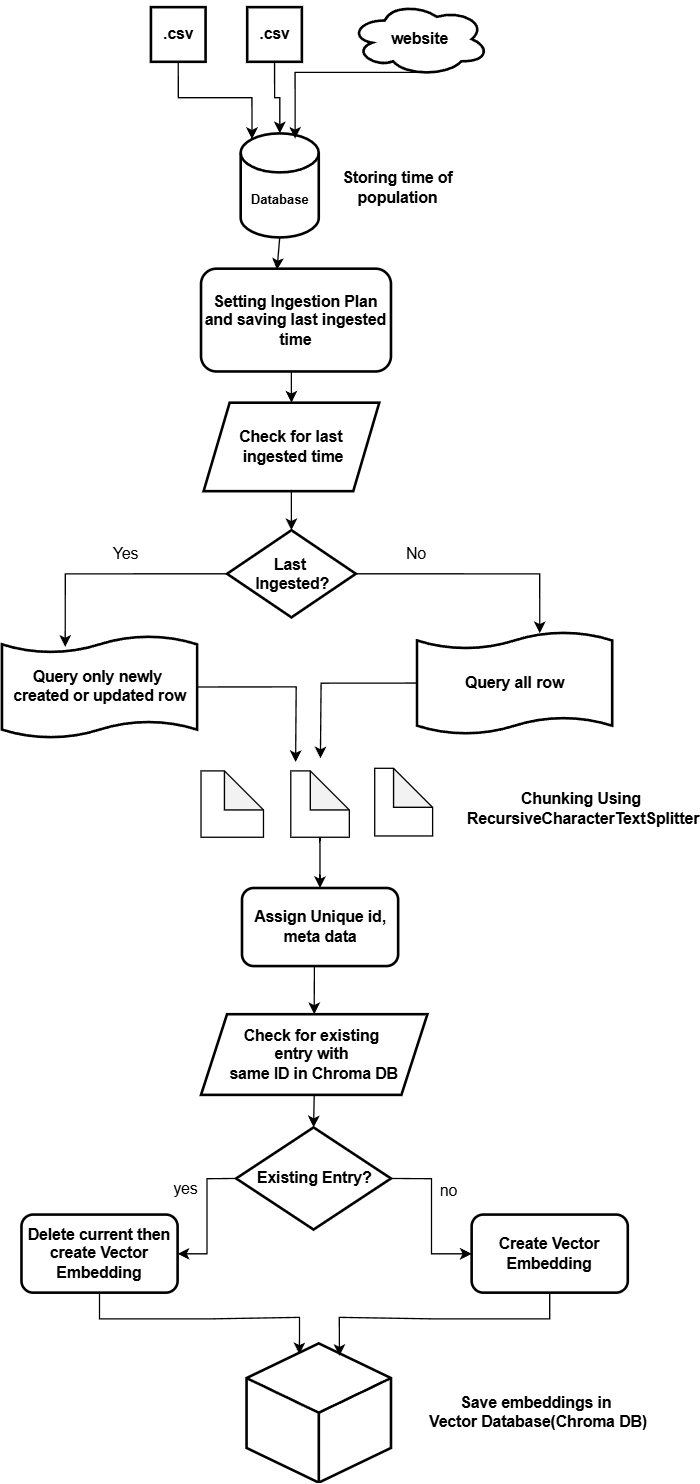} 
    \caption{Workflow of Database Population.}
    \label{fig:Workflow of Database Population}
\end{figure}

\section{MODEL}
Our chatbot is designed to assist university students by providing academic guidance through a Hybrid Retrieval Mechanism and LLM-based Response Generation. It combines BM25 lexical ranking and ChromaDB semantic retrieval to make sure that the responses are very relevant. This retrieval is complemented by LLaMA-based text generation, enhancing contextual understanding and fluency in responses.

\subsection{Hybrid Retrieval Mechanism}
This investigation uses a Hybrid Retrieval Model that integrates BM25 lexical ranking techniques with ChromaDB vector retrieval to optimize information retrieval. The approach provides correct search outcomes independent of whether users type literal terms or conceptual equivalents. Once the user sends a query, the system tokenizes it and prepares it for subsequent steps.
\subsubsection{BM25 Scoring}
The Best Matching 25~\cite{robertson2009bm25} ranking function identified as BM25 serves information retrieval systems by measuring document relevance to search queries. The Basic Matching model represents an upgraded version of TF-IDF (Term Frequency-Inverse Document Frequency) standards which function extensively throughout search engine applications. 
\subsubsection{ChromaDB Similarity}
Like BM25, ChromaDB returns the top k results but here less distance indicates more similarity. The system retrieves 10 documents that possess similar vectors. During semantic retrieval the system identifies relevant documents which demonstrate conceptual or contextual relationships with the query despite lacking direct lexical matches because such associations remain vital for discovering more abstract topic-oriented content. Again if there is any common document in top 10 of BM25 and Chroma DB then it means there is already an entry for that document in the combined docs list. So we just update the place holder for the distance key to the document’s distance. If any document of Chroma DB is absent in BM25’s result then a new entry is registered into the combined doc where the value of BM25 is set to zero.
\subsubsection{ Normalizing and Combined Scoring}

To produce a single combined score, the retrieval system must handle two very different measures—BM25 scores and vector distances: \par 

The system performs BM25 Normalization through score multiplication by the candidate set 
maximum divisor to generate values between 0 to 1. These scores are stored in the 
\textbf{bm25\_normalized} array. 
The system adjusts vector distance values so that closer correspondences transform into 
higher numerical results. The function \textbf{distance\_similarity = 1/(1+distance)} 
provides simple normalization techniques.

\par

\textbf{Weighted Combination:} The final document ranking system uses weighted sums of normalized BM25 scores with normalized vector similarities to compute its final score. For example, the combined score formula incorporates: 
{\small
\begin{equation}
\mathrm{Combined Score}(D,Q) = \lambda \cdot \mathrm{BM25}(D,Q) + (1 - \lambda) \cdot \mathrm{Similarity}(D,Q)
\end{equation}
}
\par 

Here w(BM25) and w(Chroma) are the weights that can be adjusted to customize the combined score which enables users to adjust lexical and semantic weight contributions. We have kept both the weights to 0.5 which gives equal preference to both lexical and semantic values. \par 

\subsubsection{Final Ranking}
Once the combined score for each document is calculated: \par 

\textbf{Sorting:} Documents are sorted in descending order based on combination scores, so relevant documents that are the highest in terms of both lexical and semantic evaluation rise to the top position. \par 

\textbf{Top-K Results:} As the last step, the retriever supplies the top k documents to users or downstream programs. The retrieval system provides ranked lists of documents which users or downstream applications need for data processing purposes or screen display needs. \par 

By balancing the BM25 and ChromaDB scores, the hybrid retriever capitalizes on the strengths of each method: \par 

Through BM25, the search precision remains high for queries requiring exact word matches.
Through ChromaDB users can draw out semantic information with contextual understanding which simple lexico-semantic matching often overlooks. This methodological integration delivers a strong flexible retrieval system suitable for diverse application scenarios ranging from basic keyword retrieval to deep semantic investigations.
\begin{figure}[htbp]
    \centering
    \includegraphics[width=0.40\textwidth]{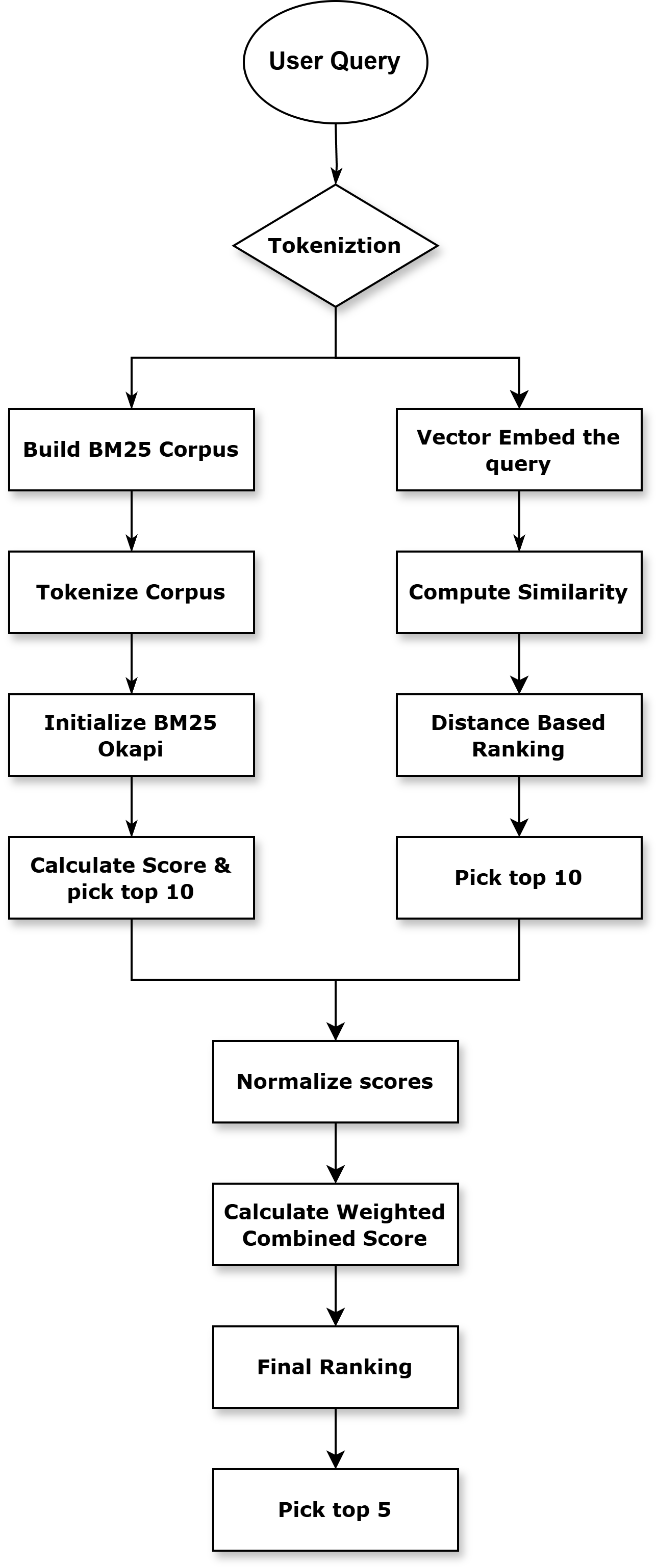} 
    \caption{Hybrid retriever workflow.}
    \label{fig:hybrid_retriever}
\end{figure}

\subsection{Response Generation}
The automated response system uses multiple stages that guarantee accurate
interactions while delivering contextually relevant and properly formatted results.
These steps include:
\subsubsection{User Query Processing}
The user query processor operates through Streamlit as part of the chatbot
interface to provide real-time query handling. The chatbot processes a user’s input
then keeps the information in its session state. The system allows the bot to track
messages history along with prior conversations which stands as a vital
requirement for contextual response generation. By storing session history, the bot
maintains seamless conversations that flow naturally instead of delivering random
or split responses.
\subsubsection{Hybrid Retrieval for Contextual Information}
The hybrid retrieval system serves to improve the chatbot's response knowledge base. The retrieval system combines keyword- and semantic-search technologies to deliver accurate results from a context-based perspective. The chatbot utilizes: \par 

\begin{itemize}
    \item \textbf{BM25 Algorithm:} The document scoring system uses a probabilistic ranking function which evaluates document relevance through term frequency determination along with inverse document frequency analysis (TF-IDF).

    \item \textbf{ChromaDB Embeddings:} Semantic search utilizes vector similarity to retrieve documents that preserve contextual meaning through its retrieval process.

\end{itemize}
\par 

The hybrid system keeps the balance of retrieval by using both keyword-based BM25 to find exact matches and embedding-based retrieval to find meanings that are related in text. The retrieval process produces rankings of relevant documents that limit response generation to top-ranked contexts. \par 

\subsubsection{Contextual Representation for Response Generation}
The system arranges retrieved documents that show the highest relevance into an
organized structure. The constructed response uses a combination of user query
and hybrid retriever context alongside corresponding references from the
knowledge base. The organized format of this representation guarantees both valid
and source-based backing for every response.The systemized method for arranging
collected data enables the chatbot to produce responses which match user queries
while maintaining factual accuracy. Additional security comes from the chatbot
referencing its source material that builds user trust together with credibility.
Users who access source references in the chatbot system can check the validity of
generated responses to enhance the accuracy of chatbot education outcomes.

\subsubsection{LLM-Based Response Generation}
Through GROQ LLaMA-3.3-70B - a modern language model - the system
develops natural language responses. Consistency and accuracy in the chatbot
initiation phase require the definition of a system message which defines the
essential foundation of an interaction by explaining how the assistant functions
while also determining response types and behavior boundaries.

\subsubsection{Response Delivery and Conversation Management}

Following its response generation process the LLM requires various post-processing
operations to create final user-ready content. Users see the response produced by
the LLaMA-3.3-70B model when it appears in the chat interface for their
interaction. The session history stores the response which maintains dialogue
coherence and creates an ongoing context for following interaction.  The controlled delivery
process preserves the chatbot’s ability to keep track of context and answer
effectively to multiple queries while respecting ethical standards in all
conversations.
\subsubsection{Error Handling and API Failure}

An exception handling system within the chatbot enables the proper management
of failures, including API outages or connectivity problems. When the LLM
encounters an unavailable state, it notifies users properly while advising them to
attempt again at a later time. Through this proactive measure, the system prevents
feedback loss in situations of unexpected failures, which ensures reliable user
experiences
\begin{figure}[htbp]
    \centering
    \includegraphics[width=0.56\textwidth]{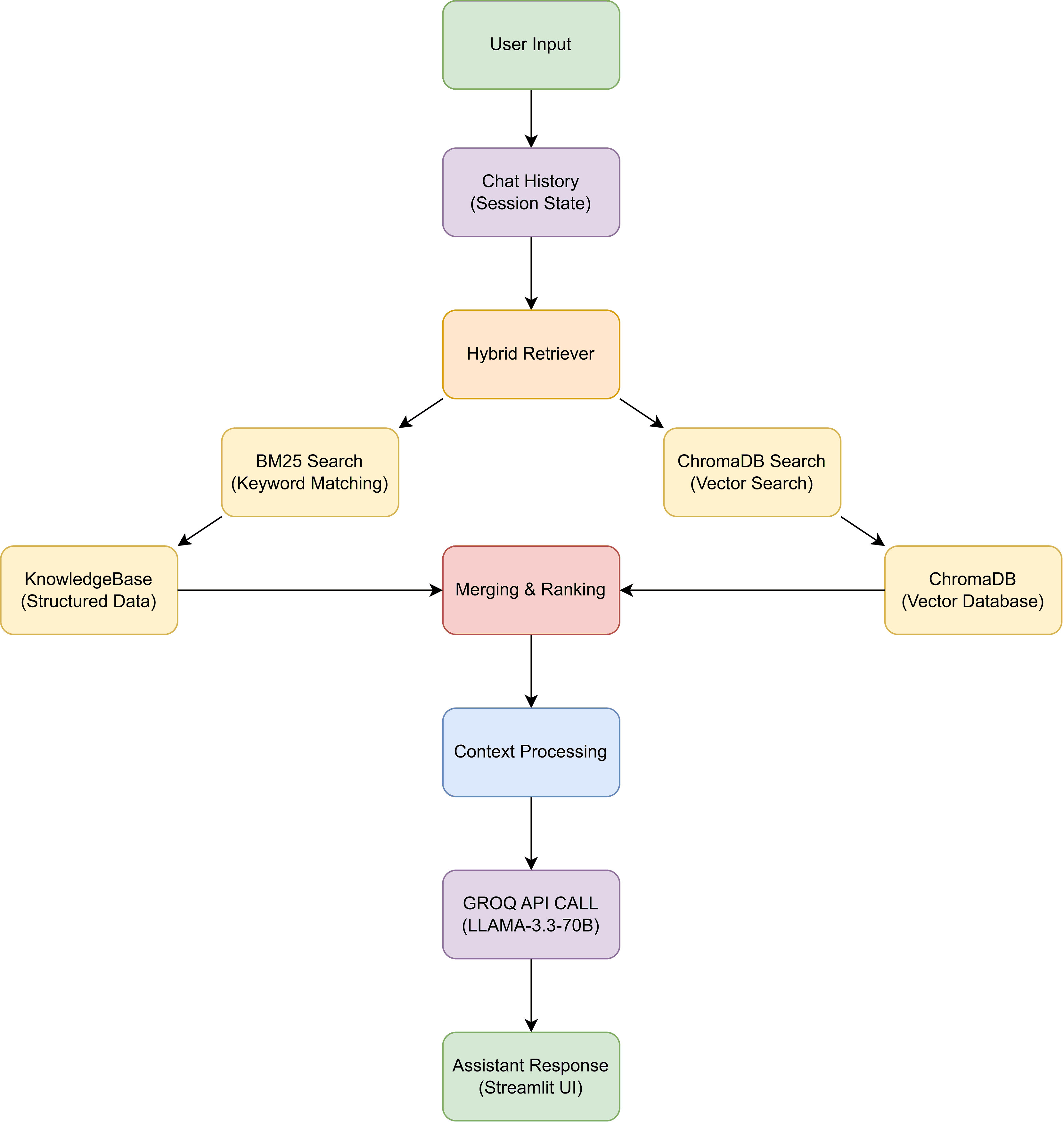} 
    \caption{Workflow of Response Generation.}
    \label{fig:response_generation}
\end{figure}
\section{Result and Analysis}
The section starts with comparing the time requirements for ingesting new data, updated data and old data. This comparison clearly shows the benefits of our data ingestion pipeline. When the knowledgebase was empty and we were ingesting total new data, the pipeline required 368.62 seconds. On contrary, updating the CSV files and then again ingesting the data took only 106.82 seconds. Lastly, we tried to ingest the old data but it took only 4.37 seconds which was required to intialize embedding model. This difference in time requirements clearly shows that our ingestion pipeline is useful for this research.

\begin{figure}[htbp]
    \centering
    \includegraphics[width=0.46\textwidth]{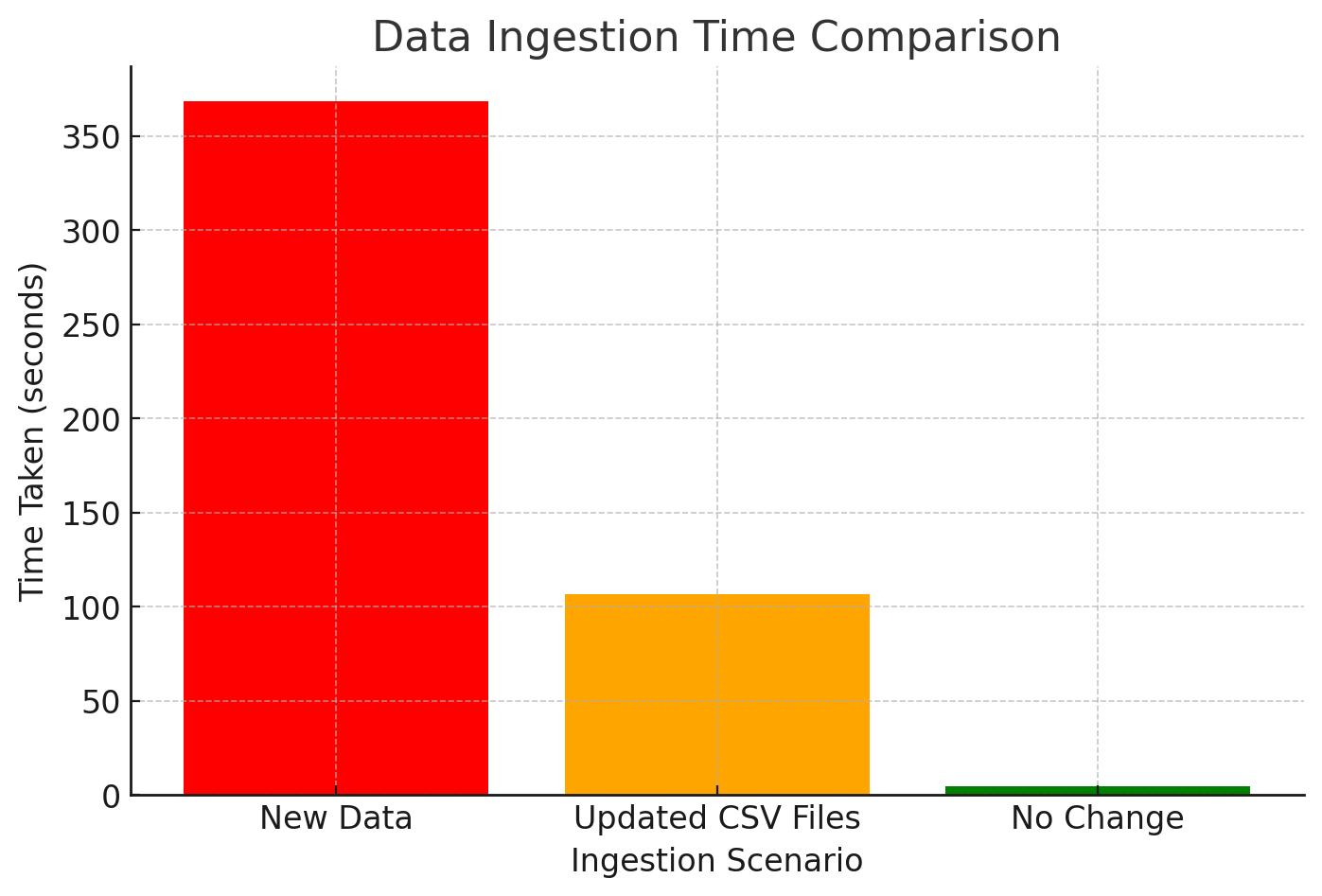} 
    \caption{Data Ingestion Time Comparison.}
    \label{fig:Data Ingestion Time Comparison}
\end{figure}

This section also offers a detailed analysis of the model’s performance in the
question-answering task. We evaluated the fine-tuned BERT model’s behavior
across different learning rates, its training and validation loss progression, and its
ability to handle queries effectively. Additionally, we used BLEU and ROUGE
metrics to provide a more comprehensive assessment of the model’s performance.
To assess the quality of the generated responses, we evaluated our model using
four widely recognized automatic metrics: BLEU, ROUGE-L, BERTScore, and
METEOR. The metrics measure distinct aspects of text generation starting from
lexical overlap moving through syntactic structure with semantic coherence.
\begin{figure}[htbp]
\centering
\includegraphics[scale=0.6]{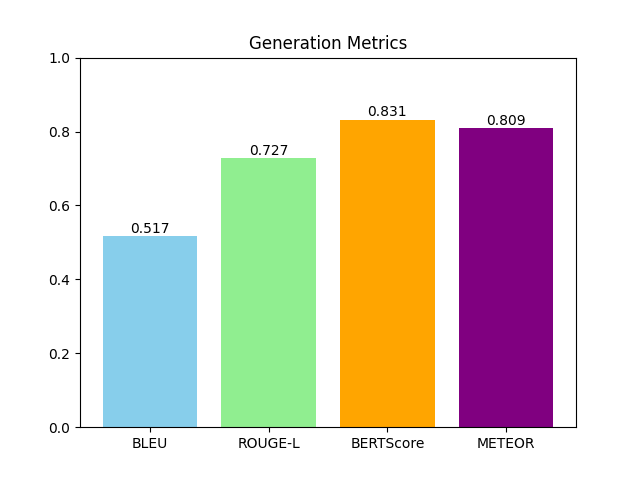}
\caption{Comparison among evaluation metrics}
\label{fig:x Comparison among evaluation metrics}
\end{figure}

\subsection{Bilingual Evaluation Understudy}
BLEU- Bilingual Evaluation Understudy score serves as the evaluation instrument for machine translation quality and natural language processing model outputs. The metric demonstrates the level of matching between a text and its reference. Under optimal alignment conditions with reference text the BLEU score reaches between 0 and 1. 

\[
\mathrm{BLEU}
= BP
\,\cdot\,
\exp\!\Bigl(\,\sum_{n=1}^{N} w_{n}\,\log p_{n}\Bigr).
\]

\[
BP = 
\begin{cases}
1, & \text{if } c > r, \\[6pt]
e^{\,\bigl(1 - \frac{r}{c}\bigr)}, & \text{if } c \le r
\end{cases} \tag{\cite{papineni2002bleu}}
\]
\par

Our model shows 0.517 performance according to the BLEU evaluation standard which calculates how well the produced text matches against references. Our model showed a BLEU score of 0.517 which demonstrates moderate word pairing alignment. BLEU shows good matching results but since it avoids semantics and synonyms it remains an unreliable tool for language generation systems.

\subsection{Recall-Oriented Understudy for Gisting Evaluation}
One of the most popular metrics for evaluating text generation models operates under the name ROUGE-L (Recall-Oriented Understudy for Gisting Evaluation - Longest Common Subsequence) and works specifically with summarization and dialogue systems. This similarity measure finds the longest common sequence of words that exist between reference and generated text outputs. The n-gram match requirement of BLEU does not measure sentence structure and fluency at the level ROUGE-L does with its LCS evaluation. \par \vspace{0.25 cm}
\[
\mathrm{ROUGE\text{-}L}
= \frac{\mathrm{LCS}(X,Y)}{\lvert Y\rvert}
\]
\[
F_{1}
= \frac{\bigl(1 + \beta^{2}\bigr)\,R\,P}{
   R \;+\; \beta^{2}\,P
} \tag{\cite{lin2004rouge}}
\]

ROUGE-L uses the longest common subsequence or LCS algorithm to measure text structure similarity between generated and reference content. The system scored 0.727 in ROUGE-L F1 which shows that its responses maintain most of the reference answers' basic structure. The results show that the model creates answers with excellent grammatical flow.

\subsection{BERTScore}
Using pre-trained transformer embeddings the BERTScore approach calculates token-level semantic similarities beyond traditional metrics BLEU and ROUGE. Within our evaluation our model achieved a BERTScore of 0.831 marking the top result among all metrics. The measurement shows that model responses keep sensible semantic connections with human references despite lacking exact word correspondence. BERTScore demonstrates its capability to understand contextual relationships because it produces natural responses which match reference texts appropriately. 

\[
\mathrm{BERTScore}
= \frac{1}{\lvert X\rvert}
\,\sum_{x \in X}
\max_{y \in Y}
\cos\!\bigl(E(x),\,E(y)\bigr) \tag{\cite{zhang2019bertscore}}
\]

\subsection{Metric for Evaluation of Translation with Explicit ORdering )} \par 
The evaluation metric METEOR (Metric for Evaluation of Translation with Explicit ORdering) serves to assess machine-generated texts especially within translation and text generation applications. The method makes BLEU better by letting users rate the quality of translations by finding synonyms, analyzing word roots, and judging the order of words, with recall accuracy being the most important factor. The ability of METEOR to extract semantic meaning and variety in language generates a precise analysis of text quality. \par 
\[
F_{\mathrm{mean}}
= \frac{P \cdot R}{
   \alpha \,P \;+\; \bigl(1 - \alpha\bigr)\,R
}.
\]
\[
\mathrm{Penalty}
= \gamma \,\Bigl(\frac{ch}{m}\Bigr)^{\theta}
\]
\[
\mathrm{METEOR}
= F_{\mathrm{mean}} \,\times\,\bigl(1 \;-\; \mathrm{Penalty}\bigr) \tag{\cite{banerjee2005meteor}}
\]

\par

The METEOR metric measures translation quality through synonymy evaluation and stemmed term analysis and word placement assessment,aking it more versatile than BLEU. Our system scored 0.809 in METEOR evaluation,indicating superior alignment to reference responses with flexibility for different word selection. The results demonstrate that the produced texts maintain semantic relevance while showing diverse linguistic patterns.
\par
\textbf{Insights and Implications:} \vspace{0.5 cm} 

The model produces contextually relevant results because its BERTScore and METEOR metrics show high semantic matching although its BLEU and ROUGE-L metrics indicate lower lexical similarity. The NLP metrics BERTScore and METEOR prove more suitable for current NLP applications including abstractive summarization and dialogue generation and paraphrasing because they focus on meaning maintenance over literal word usage. The standard n-gram quantification systems (BLEU, ROUGE) demonstrate their value in checking structural coherency yet they lack full capability to evaluate flexible language production performance of models. The findings demonstrate semantic similarity measures such as BERTScore and METEOR deliver better reliability in assessing text generation quality when compared with the traditional n-gram metrics BLEU and ROUGE-L.

\section{CONCLUSION}
\subsection{Limitation}\par
Entity resolution problems exist in the chatbot system by causing an inability to match full names with short forms thus producing different answer patterns. The problems with detecting aliases reduce system performance for information retrieval which makes output results unreliable. The restricted nature of the dataset makes its application ineffective beyond the CSE students at BRAC University since only department-specific data is included. Students who either need communication in Bangla or Banglish face difficulties accessing the chatbot because it lacks support for multiple languages. The retrieval system shows varying accuracy levels because users use different wording formats when making queries. 
\subsection{Future work}\par
Contextual consistency can improve through integration of retrieval methods such as Agentic RAG together with reinforcement learning. A knowledge graph boosts entity identification through effective linking of faculty names initial and email information. The addition of information from various departments and universities will enhance the generalizability of the dataset. Integrating multilingual LLMs mBERT and Multilingual T5 will make the chatbot more inclusive.
\subsection{Conclusion}
The chatbot utilizes BM25 together with ChromaDB-based vector retrieval to present academic and administrative information properly and efficiently to BRAC University students. The response relevance benefits from structured pipeline combined with embedding techniques which receive backing through BLEU, ROUGE-L, BERTScore and METEOR evaluations. This system encounters three significant problems which include inconsistent alias formats alongside departmental information biases and a inability to handle multiple languages. The system can become more beneficial for worldwide student use through planned developments involving reinforcement learning together with knowledge graphs and extended datasets as well as multilingual functionality. The enhanced system will transform the chatbot into an educational assistance system applicable for universities across the globe. The next stage of development will include better response customization along with more competent management of intricate inquiries to reach optimal user experience standards. The existing restrictions of the chatbot system could be resolved to establish AI academic support as an industry benchmark.

\end{document}